# Plasmonic Lattice Mode Formed by Ag Nanospheres on Silica Pillar Arrays

Xiaodan Huang, Chaogang Lou*, Hao zhang, and Didier Pribat

*Joint International Research Laboratory of Information Display and Visualization, School of Electronic Science*

*and Engineering, Southeast University, Nanjing 210096, P. R. China*

(Dated: November 23, 2017)

A method to form the plasmonic lattice mode of periodic metallic nanoparticle arrays is presented. In the arrays, each Ag nanosphere is on the top of a $SiO_2$ nanopillar which sits on a quartz substrate. The simulated results show that, in the wavelength range around the period of the arrays, the plasmonic lattice mode can be formed. The transmittance of the structure varies rapidly from the maximum to the minimum in a narrow wavelength band. This is attributed to the wavelength-dependent diffracted waves generated by the periodic arrays and the weak near-field coupling between Ag nanospheres and the quartz substrates. Increasing the height of $SiO_2$ nanopillars can weaken the near-field coupling between Ag nanospheres and the quartz substrates, and lead to the greater effects of the diffracted waves on the transmittance. As a result, the plasmonic lattice mode can be optimized. This method provides a possible way to promote the application of the plasmonic lattice mode.

*Introduction.*—Plasmonic lattice mode (PLM) is an interesting phenomenon which occurs in the periodic arrays of metallic nanoparticles [1-14]. When the periodic arrays of the metallic nanoparticles are illuminated by the light with the wavelength around the period of the arrays, the transmittance varies rapidly from the maximum to the minimum in a narrow wavelength band [5]. The unique nature can be explained by the localized surface plasmon resonances of the metallic nanoparticles and the diffraction of the light [10-14]. Some researches have demonstrated that the remarkable optical property offers the opportunities for developing new sensors or lasers [15-17].

However, applying PLM in devices faces a challenge. The formation of PLM requires that the periodic arrays of the metallic nanoparticles should be surrounded by homogeneous (or index-matched) materials because inhomogeneous surroundings contort the diffraction and suppress the formation of PLM [18-20]. Some works have reported that, when the metallic nanoparticles are deposited on a substrate, PLM could not be observed [19]. This largely limits the applications of PLM because in actual devices the metallic nanoparticles are usually located on the substrate whose refractive index is different from superstrates.

Here an approach is proposed to form PLM without homogeneous surroundings. In the method, $SiO_2$ nanopillars are introduced between a periodic Ag nanosphere array and a quartz substrate. Each Ag nanosphere is on the top of each $SiO_2$ nanopillar which sits on the quartz substrate. Because the equivalent index of $SiO_2$ nanopillars is smaller than that of the quartz substrate, this brings a smaller index difference between the superstrate and the substrate of Ag nanospheres. When the periodic arrays is illuminated by light, it is possible to form PLM. The research work is carried out by simulation, and the mechanism of forming PLM is discussed.

*Modeling.*—Figure 1 shows the simulated structure. Both periods of the arrays in x and y direction are 450 nm. Ag nanospheres and $SiO_2$ nanopillars have the same diameter of 60 nm. The percentage of the substrate's surface area covered by the arrays is 1.4%. An incident light propagates along z direction. The refractive index of Ag are taken from palik [21]. In the range from 400 nm to 550 nm, the real part of Ag's refractive index varies from 0.173 to 0.125, and the imaginary part varies from 1.95 to 3.35. The refractive index of the quartz and $SiO_2$ are set as a constant value of 1.5 [21] . The simulation is carried out by using finite element method. The structure of the periodic Ag sphere arrays directly sitting on the quartz substrate without $SiO_2$ pillars is also simulated for the purpose of comparison.

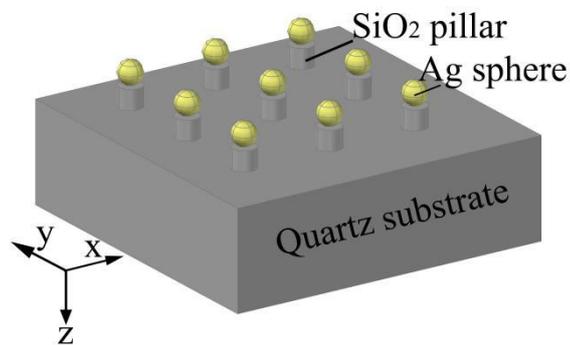

FIG. 1. The schematic of the periodic arrays with Ag nanospheres and $SiO_2$ nanopillars on a quartz substrate.

*Results.*—The red curve in Fig. 2 shows the transmittance of the simulated structure with $SiO_2$ pillars of height 300 nm. It can be seen that the transmittance increases with the wavelength until it reaches the maximum around 450 nm. Once the wavelength becomes longer than 450 nm, the transmittance drops rapidly down to the minimum around the wavelength 478 nm. After that, the transmittance turns to rise again. This is the typical PLM of the periodic metallic nanoparticles [1,2]. While, for the structure without $SiO_2$ pillars, PLM does not appear, and the transmittance (black curve) rises gradually with the wavelength.

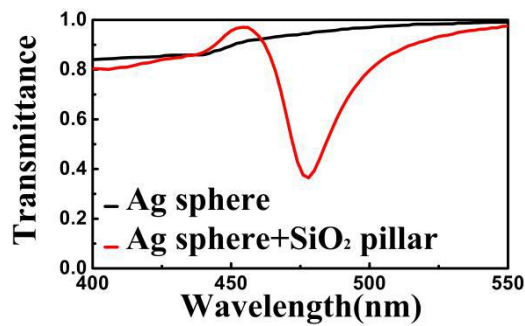

FIG. 2. The transmittance of the simulated structures. The red curve is for Ag spheres on $SiO_2$ pillars (of height 300 nm) which sit on the quartz substrate. The black curve is for Ag spheres directly sitting on the quartz substrate.

Although the transmittance is mainly determined by the directly transmitted light which is not scattered by Ag sphere arrays because the coverage of the arrays is as low as 1.4%, the two curves in Fig. 2 indicate that the arrays have important effects on the transmittance. The only difference between the two structures is $SiO_2$ pillars, so it is natural to think that $SiO_2$ pillars lead to the different variations of the two curves. To explain the phenomenon, we should start from the diffraction of the periodic arrays and the near-field coupling between Ag spheres and the substrate.

According to the diffraction theory [22,23], when the incident wavelength is equal to 450 nm (the period of the arrays), there are only two orders of the diffracted waves: the zero-order propagates vertically and the first-order propagates horizontally, as shown in Fig. 3 [22]. For the simulated structure in Fig. 1, if the incident wavelength is shorter than 450 nm, the propagating direction of the zero-order waves is still vertical, but the first-order waves propagate along the direction which has an angle with the horizontal direction. When the incident wavelength is longer than 450 nm, the first-order diffracted waves disappear, and the zero-order diffracted waves still propagate vertically. For the periodic Ag sphere arrays, localized surface plasmonic resonance is formed when they are illuminated by the incident light. If the incident wavelength is around the period of the arrays, the near-field energy around Ag spheres will be dissipated through the diffracted waves, and consequently the intensity of the near-field is also changed [23,24].

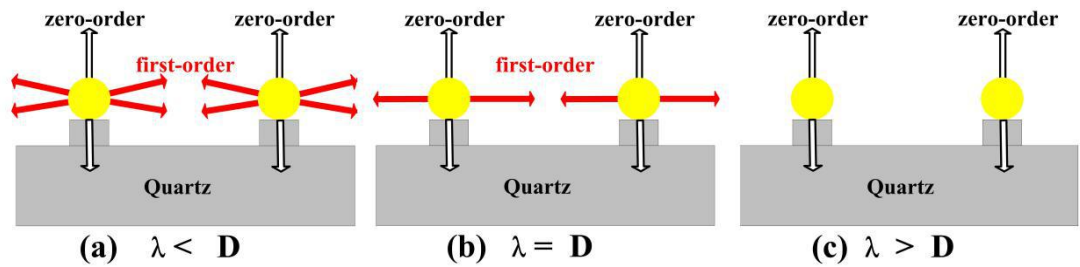

FIG. 3. Diffracted waves of the structure under different incident wavelengths. λ is the incident wavelength and D is the period of the arrays.

Besides the diffraction, the near-field coupling between Ag spheres and the substrate also has effects on the near-field intensity because it transmits the near-field energy into the substrate and weaken the near field. The effects depend on the distance between Ag spheres and the substrate because the near field drops down quickly with the distance. For the periodic arrays with higher $SiO_2$ pillars shown in Fig. 1, the near-field coupling between Ag spheres and the substrate is weak, and have less effects on the near-field intensity of Ag spheres than the diffraction.

The periodic arrays influence the transmittance through the extinction cross-section of Ag spheres. The bigger extinction cross-section results in lower transmittance. Different from a single isolated Ag sphere on a 300 nm $SiO_2$ pillar whose extinction cross-section reaches maximum at its resonance wavelength, the extinction cross-section of the periodic Ag sphere arrays reaches the maximum at the wavelength where the near field is the strongest. This wavelength is different from the resonance wavelength of the single isolated Ag sphere and depends on the diffraction.

Figure 4 gives the extinction cross-sections of the single Ag sphere on a 300nm $SiO_2$ pillar and the periodic arrays of Ag spheres on 300 nm SiO$_2$ pillars. It can be seen that the extinction cross-section of the periodic arrays reaches the minimum around 450 nm and the maximum around 478 nm, while the single Ag sphere reaches the maximum extinction cross-section around 440 nm which is the resonance wavelength of the single Ag sphere.

When the incident wavelength rises from 440 nm to 450 nm, the diffracted waves become stronger because the wavelength is approaching the period of the structure [23,24]. Different from the isolated Ag spheres around which the near-field energy is concentrated, the near-field energy of the periodic Ag sphere arrays can be dissipated through the diffraction waves [23,24]. This weakens the intensity of near field and the polarization of Ag spheres. Because the extinction across-section of Ag spheres depends on their polarization, when the incident wavelength rises, the extinction across-section decreases and results in the increasing transmittance [5].

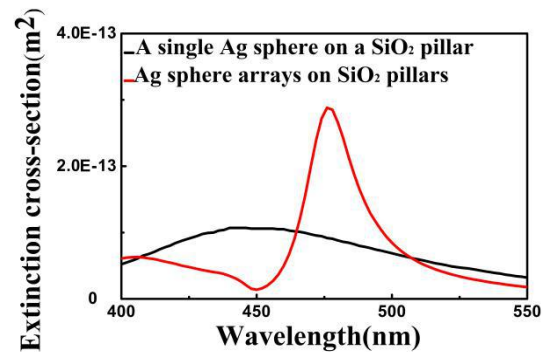

FIG. 4. The calculated extinction cross-section for different structures. The black curve is for an isolated Ag sphere on a SiO$_2$ pillar. The red curve is for Ag sphere arrays on SiO$_2$ pillars.

When the incident wavelength continues to rise beyond 450 nm, the first-order diffracted waves disappear rapidly [23,24]. The near-field energy can no longer dissipate through the first-order diffracted waves and is concentrated around Ag spheres. This strengthens their driving electric field and results in the quick increase of the extinction across-section until the the first-order diffracted waves disappear completely (shown in Fig. 4). On the other hand, when the incident wavelength is leaving away from the resonance wavelength 440 nm, the resonance of Ag spheres is weakened. This will cause the decreased extinction across-section. The competition between these two mechanisms results in the maximum of the extinction cross-section of Ag spheres and the minimal transmittance around 478 nm. After that, the weakened resonance becomes dominant in determining the extinction cross-section, so the transmittance turns to increase.

However, for the structure which does not include SiO$_2$ pillars, Ag spheres sit directly on the quartz substrate. The near-field coupling between Ag spheres and the quartz substrate become strong and important to the transmittance. In the wavelength range of 440-450 nm, the diffraction is strengthened with the rising wavelength, and most of the near-field energy is dissipated through the diffracted waves, so the electric field around Ag spheres is weak, and the transmittance increases with the wavelength due to the decreased extinction cross-section which is caused by weakened polarization.

After the incident wavelength becomes longer than 450 nm, the first-order diffracted waves disappear, so the dissipated energy through the diffraction decreases and leads to a strengthened near field which enlarges the extinction cross-section and lowers the transmittance. On the other hand, because Ag spheres sit directly on the substrate, the strengthened near-field coupling between Ag spheres and the substrate transmits the more near-field energy into the substrate [25]. This offsets the effect of the disappeared first-order diffracted waves on the extinction cross-section and enhances the transmittance. Considering that the near-field coupling can also increase the transmittance, the total transmittance keeps rising when the wavelength is beyond 450 nm, as shown in Fig. 2.

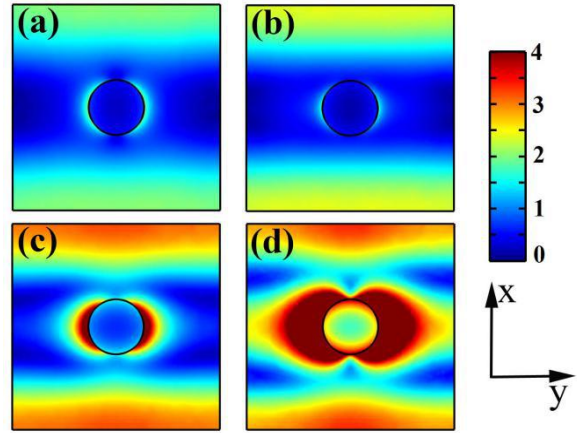

FIG. 5. The local electric distribution around Ag spheres on $SiO_2$ pillars of height 300 nm at different wavelengths. (a) 440nm, (b) 450nm, (c) 460nm and (d) 478nm. The electric field is normalized by the incident field intensity. The plots are in x-y plane of the arrays.

Figure 5 shows the local electric field distribution around Ag spheres on $SiO_2$ pillars of height 300 nm at different wavelengths. It can be seen that the local electric field around Ag spheres is weakened from 440 nm to 450 nm, and then become stronger at 460 nm. At the wavelength 478 nm, the local field reaches the maximum. This agrees with the explanation about the extinction cross-section and the transmittance.

Figure 6(a) shows the transmittance of Ag spheres on $SiO_2$ pillars of different heights. With the increasing pillars' height, the dip of the transmittance drops down. This might be attributed to the near-field coupling between Ag spheres and the quartz substrate, which is weakened with the rising pillars' height. In the wavelength range above 450 nm, the first-order diffracted waves disappear, so the incident energy can not be dissipated through the first-order diffracted waves and forms the near field around Ag spheres. In this case, the higher $SiO_2$ pillar results in the weaker near-field coupling which leads to the higher electric field around Ag spheres and the bigger extinction cross-section, so the dip of the transmittance is lowered.

Figure 6(b) shows the relationship between the transmittance at the dip's position and $SiO_2$ pillars' height. When the height of $SiO_2$ pillars increases from 10 nm to 1400 nm, the transmittance decreases due to the weakened near-field coupling between Ag spheres and the substrate. Especially, for the pillar's height shorter than 500 nm, the transmittance decreases rapidly. This is because the near-field coupling decreases rapidly for the lower pillars' height and slowly for higher pillars' height.

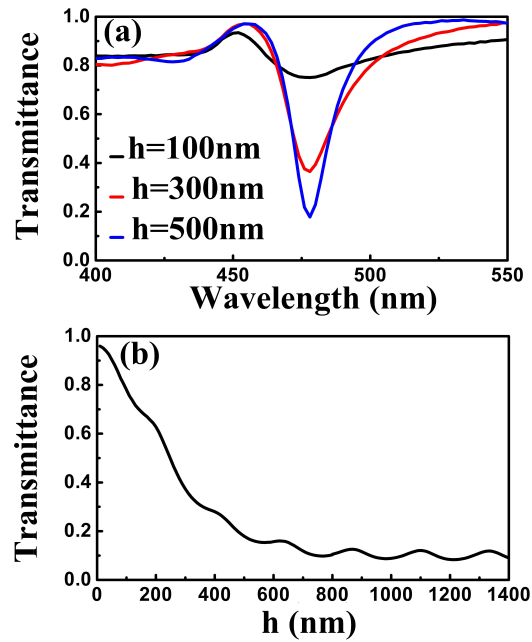

FIG. 6. (a) The calculated transmittance of Ag spheres on $SiO_2$ pillars with different $SiO_2$ pillars' heights of 100nm, 300nm, 500nm. (b) The variation of the transmittance's dip with $SiO_2$ pillars of different heights which vary from 10nm to 1400nm.

*Conclusion.—*The plasmonic lattice mode is demonstrated in the structure with Ag spheres on $SiO_2$ pillar arrays which sit on a quartz substrate. The introduction of $SiO_2$ pillars weakens the near-field coupling between Ag spheres and the substrate, and makes the transmittance vary rapidly from the maximum to the minimum in the wavelength range 450-480 nm. By increasing the height of the $SiO_2$ pillars, the transmittance at the dip's position can be lowered and the plasmonic lattice mode can be optimized. This provides a possible way to promote the application of the plasmonic lattice mode.

The authors thank the supports from the Natural Science Foundation of Jiangsu (Grant No. BK2011033) and the Primary Research & Development Plan of Jiangsu Province (Grant No. BE2016175).

*Electronic address: *lcg@seu.edu.cn